\documentstyle[12pt]{article}

\newlength{\dinwidth}
\newlength{\dinmargin}
\def\lapproxeq{\lower .7ex\hbox{$\;\stackrel{\textstyle
<}{\sim}\;$}}
\def\gapproxeq{\lower .7ex\hbox{$\;\stackrel{\textstyle
>}{\sim}\;$}}

\begin{document}
\titlepage
\begin{flushright}
DTP/95/96  \\
CBPF-NF-079/95 \\
RAL-TR-95-065\\
November 1995 \\
revised February 1996\\
\end{flushright}

\begin{center}
\vspace*{2cm}
{\Large {\bf Diffractive $J/\psi$ photoproduction as a probe of
the gluon density}}
\\
\vspace*{1cm}
M.\ G.\ Ryskin$^a$, R.\ G.\ Roberts$^b$, A.\ D.\ Martin$^c$ and
E.\ M.\ Levin$^{a,d}$, \\

\vspace*{0.3cm}
$^a$  Petersburg Nuclear Physics Institute, 188350, Gatchina,
St.\ Petersburg, Russia. \\

$^b$  Rutherford Appleton Laboratory, Chilton, OX11 0QX, UK.
\\

$^c$  Department of Physics, University of Durham, Durham, DH1
3LE, UK. \\

$^d$  LAFEX, Centro Brasileiro de Pesquisas Fisicas, 22290--180,
Rio de Janeiro, Brazil. \\

\end{center}

\vspace*{3cm}
\begin{abstract}
We use perturbative QCD, beyond the leading $\ln Q^2$
approximation, to show how measurements of diffractive $J/\psi$
production at HERA can provide a sensitive probe of the gluon
density of the proton at small values of Bjorken $x$.  We
estimate both the effect of the relativistic motion of the $c$
and $\overline{c}$ within the $J/\psi$ and of the rescattering of
the $c\overline{c}$ quark pair on the proton.  We find that the
available data for diffractive $J/\psi$ photoproduction can
discriminate between the gluon distributions of the most recent
sets of partons.
\end{abstract}

\newpage

\noindent {\large \bf 1.  Introduction}

The observation of high energy diffractive $J/\psi$ photo- or
electroproduction, $\gamma^{(*)} p \rightarrow J/\psi \: p$,
offers a unique opportunity to measure the gluon density in the
proton at low $x$.  Indeed, for sufficiently high $\gamma
p$ centre-of-mass energy $W$, perturbative QCD can be used to
express the cross section for this, essentially
elastic\footnote{Elastic in the sense that the photon and
$J/\psi$ have the same quantum numbers.}, process in
terms of the {\it square} of the gluon density.  To
lowest order the $\gamma^* p \rightarrow J/\psi \: p$ amplitude
can be factored into the product of the $\gamma \rightarrow
c\overline{c}$ transition, the scattering of the $c\overline{c}$
system on the proton via (colourless) two-gluon exchange, and
finally the formation of the $J/\psi$ from the outgoing
$c\overline{c}$ pair.  The crucial observation is that at high
$W$ the scattering on the proton occurs over a much shorter
timescale than the $\gamma \rightarrow c\overline{c}$
fluctuation or the $J/\psi$ formation times, see Fig.\ 1.
\begin{figure}[htb]
\vspace{6cm}
\includegraphics{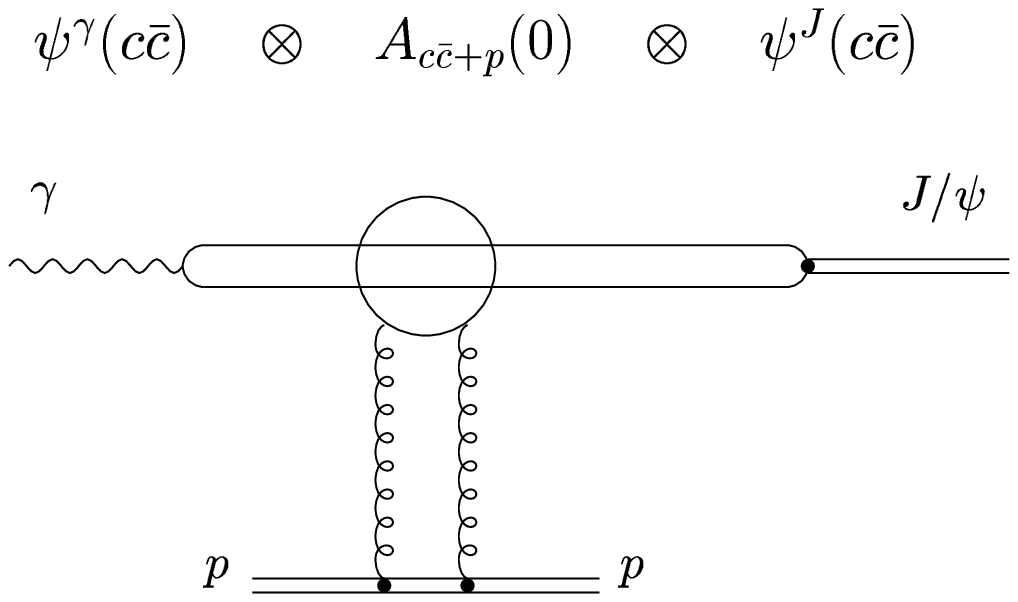}
\caption{ {\normalsize Schematic diagram for high energy
diffractive
$J/\psi$ photoproduction.  The factorized form follows since, in
the proton rest frame, the formation time $\tau_f \simeq 2
E_\gamma/(Q^2 + M_\psi^2)$ is much greater than the interaction
time $\tau_{\rm int} \simeq R$ where $R$ is the radius of the
proton.}}
\label{fig:1}
\end{figure}

\noindent Moreover this two-gluon exchange amplitude can be shown
to be
directly proportional to the gluon density $x g (x,
\overline{Q}^2)$ with
\begin{equation}
\overline{Q}^2 \; = \; (Q^2 + M_\psi^2)/4, \;\;\;\;\;\; x \; = \;
4 \overline{Q}^2/W^2.
\label{eq:a1}
\end{equation}
\noindent $Q^2$ is the virtuality of the photon and $M_\psi$ is
the rest mass of the $J/\psi$.  To be explicit, the lowest-order
formula is\cite{ry}
\begin{equation}
\left . \frac{d \sigma}{d t} \; (\gamma^* p \rightarrow \psi p)
\right |_0 \; = \; \frac{\Gamma_{ee} \: M_\psi^3 \: \pi^3}{48
\alpha} \; \frac{\alpha_S (\overline{Q}^2)^2}{\overline{Q}^8} \;
\left [ x g (x, \overline{Q}^2) \right ]^2 \; \left ( 1 +
\frac{Q^2}{M_\psi^2} \right ).
\label{eq:a2}
\end{equation}
\noindent The derivation of the result is sketched in section
2(a).  An analogous formula to (\ref{eq:a2}) was presented by
Brodsky et al.\ \cite{brod} but only for longitudinally polarized
vector mesons\footnote{It has been pointed out \cite{chan} that a
factor of 4 should be included in the numerator of the formula in
ref.\ \cite{brod}.  Then the results of refs.\ \cite{ry} and
\cite{brod} agree.}.  An earlier work \cite{dl} presented a
\lq\lq soft" Pomeron treatment of diffractive vector meson
production, but in this case the connection to the gluon
distribution $g (x, \overline{Q}^2)$ was not made.  Here we note
that the heavy meson mass $M_\psi$ should ensure that
perturbative QCD can be applied even in the photoproduction
limit, $Q^2 = 0$.  The last term in (\ref{eq:a2}) allows for
electroproduction via longitudinally polarised virtual photons
with $\sigma_L/\sigma_T \approx Q^2/M_\psi^2$.  The result
(\ref{eq:a2}) assumes a non-relativistic wave function for the
$J/\psi$ with the $c$ and $\overline{c}$ having momenta
$\frac{1}{2} q^J$.  Eq.\ (\ref{eq:a2}) is derived assuming the
leading $\ln \overline{Q}^2$ approximation in the integral $d^4
k$ over the gluon loop in Fig.\ 1.  When, in section 2, we
improve on this approximation we find from the explicit form of
the integral that typically $k_T^2 \sim \overline{Q}^2$.  It is
this which specifies the scale of $\alpha_S$ in (\ref{eq:a2}).

We may use the available measurements of the $\gamma p
\rightarrow J/\psi \: p$ production cross section (together with
the observed $J/\psi$ diffractive slope $b = 4.5$ GeV$^{-2}$)
to give a first estimate the of gluon density.  The results are
shown in Fig.\ 2.  

\begin{figure}[htb]
\includegraphics{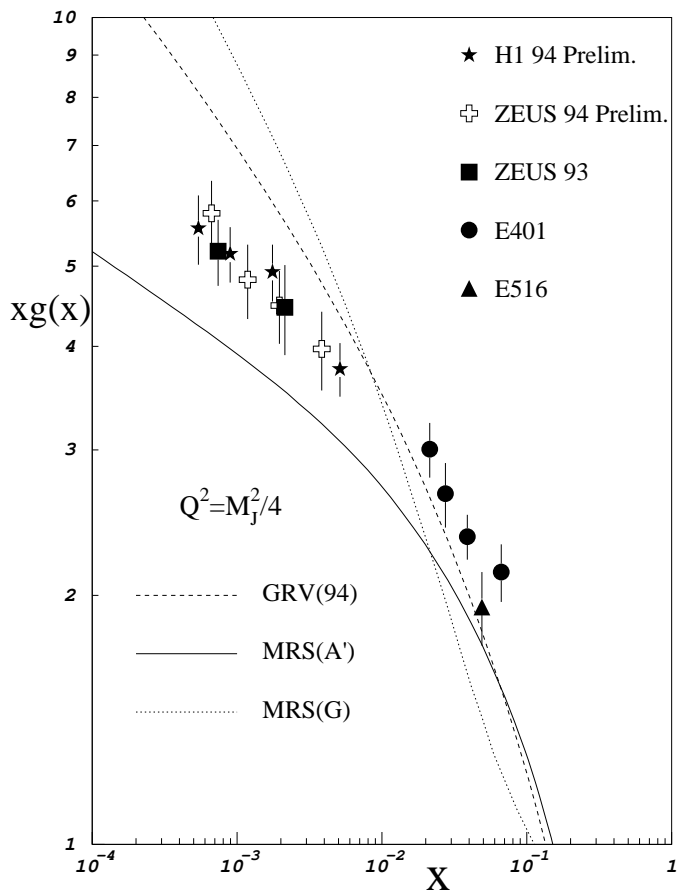}
\vspace{12.5cm} 
\caption{ {\normalsize The points show the lowest-order estimate
of the gluon density obtained from the available high energy
$\sigma (\gamma p \rightarrow J/\psi \;p)$ data [5,6] using eq.\
(2).  For comparison we also show the gluon of the GRV
parton set [7], and of the MRS(A$^\prime$, G) partons of
ref.[8] evolved back to $Q^2$ = 2.5 GeV$^2$.}}
\label{fig:2}
\end{figure}

\newpage
\noindent This figure is simply to illustrate the
typical precision and the kinematic range of the gluon density
that is probed by the $J/\psi$ data.  We emphasize that it is a
first look at the gluon and that the errors reflect only those of
experiment.  The curves correspond to gluons obtained from the
latest sets of parton distributions.  The spread of the
predictions demonstrates the potential value of this process as a
measure of the gluon.  We will show that a comparison of the {\it
shape} of the curves with the data is more reliable than that of
the {\it normalisation}.  Thus, at this preliminary stage, we
see that the shape of the $J/\psi$ data favour the
MRS(A$^\prime$) gluon.  The purpose of this paper is to refine
the theory so that a meaningful probe of the gluon distribution
can be obtained.  However, it is already clear that diffractive
$J/\psi$ production at HERA will offer a unique, precise probe of
the gluon, $xg \: (x, \overline{Q}^2)$, in the critical low $x$,
$\overline{Q}^2 \gapproxeq 2.5$ GeV$^2$ region, {\it provided}
that we can improve on the validity of (\ref{eq:a2}).

In section 2 we scrutinise the approximations used to derive
(\ref{eq:a2}) and, more important, we implement corrections so
that quantitative information on the gluon density can be
obtained.  Then, in section 3, we illustrate the discriminatory
power of the diffractive $J/\psi$ production data by comparing
with the predictions of the cross section obtained from the gluon
densities of the latest parton sets.

The gluon density at low $x$ has so far been constrained by the
slope of the deep inelastic structure function $\partial
F_2/\partial \ln Q^2$.  We find that $J/\psi$ photoproduction is
a much more sensitive measure.  This is only to be expected since
the $J/\psi$ photoproduction cross section depends quadratically
on the gluon $(\sigma \sim g^2)$, whereas in deep inelastic
scattering we have a linear dependence and then only on the
derivative of a cross section $(d \sigma \sim g)$. 

\vspace{0.25cm}
\noindent {\large \bf 2.  Improved formula for diffractive
$J/\psi$ production}
\vspace{0.5cm}

\begin{figure}[htb]
\vspace{5cm}
\includegraphics{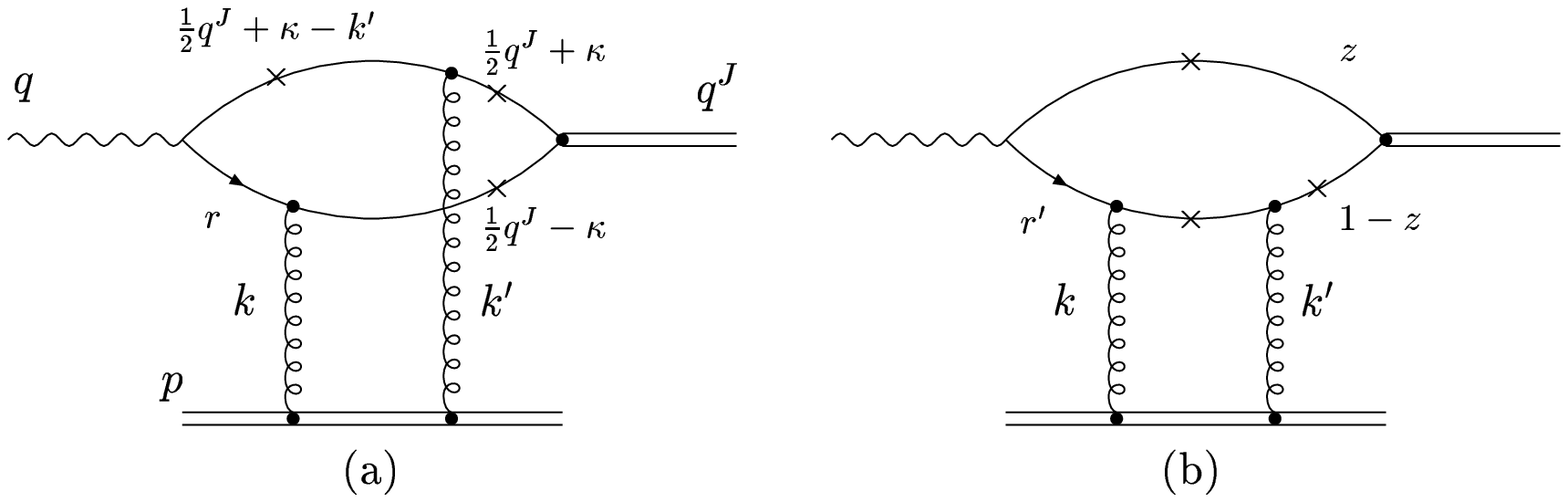}
\caption{ {\normalsize  Lowest-order perturbative QCD diagrams
for diffractive $J/\psi$ production.  The particle four-momenta
are shown; $z$ and $1-z$ denote the fractions of the longitudinal
momentum of the photon that are carried by the $c$,
$\overline{c}$ quarks.  The small crosses indicate which quarks
may be regarded as essentially on-mass-shell. }}
\label{fig:3}
\end{figure}
The amplitude for $\gamma^{(*)} p \rightarrow J/\psi \: p$ is
obtained, in lowest-order perturbative QCD, from the sum of the
two diagrams shown in Fig.\ 3. 
Formula (\ref{eq:a2}) is based on the leading $\ln
Q^2$ approximation in which we assume that the gluon transverse
momenta $k_T$ satisfies $k_T^2 \ll \overline{Q}^2$.  Moreover the
non-relativistic form is taken for the $J/\psi$ wave function 
\begin{equation}
\psi^J (z, \kappa_T) \; = \; \delta^{(2)} (\kappa_T) \; \delta
\left (z \: - \: \textstyle \frac{1}{2} \right )
\label{eq:a3}
\end{equation}
\noindent where $z = \frac{1}{2} +
\kappa_\parallel/q_\parallel^\gamma$.  
In other words, the $c$ and $\bar c$ quarks are not allowed to
have
any Fermi momentum inside the $J/\psi$, i.e. $\vec \kappa$ is set
to zero.  We discuss these, and
other, approximations in turn below.  In particular we compute
the corrections which should be applied before confronting
(\ref{eq:a2}) with the data.

\medskip
\medskip

\noindent {\bf (a) Beyond the leading $\ln Q^2$ approximation;
inclusion of the gluon $k_T$}\\

Since we are primarily concerned with diffractive $J/\psi$
photoproduction at small $x$ we work in the (leading order) $\ln
1/x$ approximation, and retain the full $Q^2$ dependence and not
just the leading $\ln Q^2$ component.  We must therefore express
the cross section in terms of an integral over the (square of
the) unintegrated gluon distribution $f (x, k_T^2)$, and so
retain the explicit gluon $k_T$ dependence.

We first evaluate the gluon loops in the Feynman diagrams shown
in Fig.\ 3.  It is convenient to perform the loop integrations in
terms of Sudakov variables.  That is the various particle four
momenta are decomposed in the form
\begin{equation}
k_i \; = \; \alpha_i q^\prime \: + \: \beta_i p^\prime \: + \:
\vec{k}_{i T}
\label{eq:z1}
\end{equation}
\noindent where $p^\prime$ and $q^\prime$ are the proton and
photon light-like momenta:  $p^{\prime 2} = q^{\prime 2} = 0$ and
$W^2 = 2 p^\prime . q^\prime$.  In particular
\begin{equation}
p \; = \; p^\prime \: + \: \alpha_p q^\prime, \;\;\;\;\;\;\; q \;
= \; q^\prime \: + \: \beta_\gamma p^\prime
\label{eq:z2}
\end{equation}
\noindent with $\alpha_p = m_p^2/W^2$ and $\beta_\gamma = -
Q^2/W^2$.  Within the non-relativistic approximation the quarks
of momenta $\frac{1}{2} q^J \pm \kappa$ are almost on-mass-shell.

The integration over the gluon longitudinal momentum puts, in the
first diagram, the upper quark with momentum $h = \frac{1}{2} q^J
+ \kappa - k^\prime$ on-shell, leaving only the quark propagator
$(r^2 - m_c^2)^{- 1}$ to be integrated over in the gluon $k_T$
integration.  To express the propagator in terms of $k_T^2$ we
first note that
\begin{equation}
r^2 \; = \; (q - h)^2 \; = \; q^2 - 2 q . h \: + \: m_c^2.
\label{eq:z3}
\end{equation}
\noindent Using the Sudakov decomposition
\begin{equation}
h \; = \; \alpha_h q^\prime \: + \: \beta_h p^\prime \: - \:
k_T^\prime
\label{eq:z4}
\end{equation}
\noindent we obtain
\begin{equation}
2 q . h \; = \; (\beta_\gamma \alpha_h \: + \: \beta_h) W^2 \; =
\; - z Q^2 \: + \: (m_c^2 \: + \: k_T^{\prime 2})/z,
\label{eq:z5}
\end{equation}
\noindent where we have neglected $\kappa_T$.  Moreover in the
non-relativistic approximation $z = \frac{1}{2}$ and so from
(\ref{eq:z3}) and (\ref{eq:z5}) we find
\begin{equation}
r^2 \: - m_c^2 \; = \; - 2 \overline{Q}^2 \: - \: 2 k_T^2 
\label{eq:z6}
\end{equation}
\noindent where $\overline{Q}^2 = (Q^2 + M_\psi^2)/4$, with $Q^2
= -q^2$, as usual.  Here we take $M_\psi^2 \simeq 4 m_c^2$.  In
Fig.\ 3(b) only the quark with momentum $r^\prime$ is off-shell. 
In analogy to the derivation of (\ref{eq:z6}) we find
$$
r^{\prime 2} \: - \: m_c^2 \; = \; - 2 \overline{Q}^2.
$$
\indent Thus the forward scattering amplitude for diffractive
$J/\psi$ production from transversely polarized photon is
\begin{equation}
A \; = \; i \: 4 \pi^2 \: M_\psi \: \alpha_S \; \int \;
\frac{dk_T^2}{k_T^4} \; \left ( \frac{1}{2 \overline{Q}^2} \: -
\: \frac{1}{2 \overline{Q}^2 + 2 k_T^2} \right ) \; G(k) \: e_c
g_J
\label{eq:a5}
\end{equation}
\noindent where colour factors give rise to the opposite sign of
the two diagrams, and where the amplitude is defined by
$$
\frac{d \sigma}{dt} \; (\gamma_T p \rightarrow J/\psi p) \; = \;
\frac{| A |^2}{16 \pi}.
$$
\noindent The constant $g_J$ specifies the $c\overline{c}$
coupling to the $J/\psi$ and $e_c$ is the charge of the $c$
quark.  The coupling $g_J$ may be determined from the width
$\Gamma_{ee}$ of the $J/\psi \rightarrow e^+ e^-$ decay, 
a process which is
described by the same $c\overline{c}$ quark loop structure.  We
have
$$
e_c^2 g_J^2 \; = \; \frac{\Gamma_{ee} M_\psi}{12 \alpha}
$$
\noindent where $e_c^2 = \frac{4}{9} \: 4 \pi \alpha$.  The
function $G(k)$ specifies the probability of finding the gluons
in the proton.  In the simplest three valence quark model
\begin{equation}
G \; = \; \frac{4}{3} \; \frac{\alpha_S}{\pi} \; 3
\label{eq:a7}
\end{equation}
\noindent where $\frac{4}{3}$ is the colour factor.

In the realistic case
\begin{equation}
G (k) \; = \; f_{\rm BFKL} \: (x, k_T^2)
\label{eq:a8}
\end{equation}
\noindent where $f_{\rm BFKL}$ is the gluon density unintegrated
over $k_T$ that satisfies the BFKL equation, an equation which
effectively resums the leading $\alpha_S \ln (1/x)$
contributions.  To gain insight into this identification, and to
specify the value of $x$, it is sufficient to study the one-rung
contribution sketched in Fig.\ 4.  

\begin{figure}[htb]
\vspace{8cm}
\includegraphics{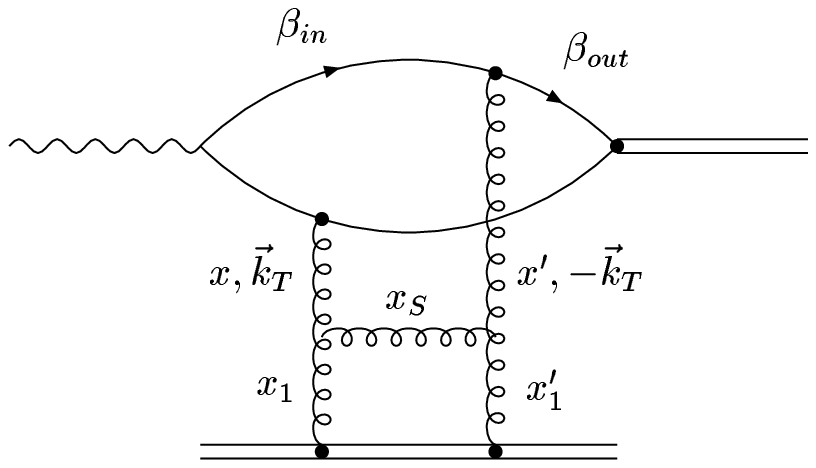}
\vspace{-3cm}
\caption{ {\normalsize  One of the one-rung diagrams which give
the
$\alpha_S \ln (1/x)$ contribution to the BFKL gluon, $f_{\rm
BFKL} (x, k_T^2)$.  The emitted gluons have proton momentum
fractions $x$ and $x^\prime$ with $|x^\prime| \ll x$.}}
\label{fig:4}
\end{figure}

To determine the momentum fractions $x$ and $x^\prime$ of the
gluons in Fig.\ 4 we note that
\begin{eqnarray}
\label{eq:b10}
x \: + \: x^\prime & = & \beta_\psi \: - \: \beta_\gamma \; = \;
\frac{Q^2 + M_\psi^2}{W^2} \\
\label{eq:c10}
x^\prime & = & \beta_{\rm out} \: - \: \beta_{\rm in} \; = \;
\frac{k_T^2}{W^2 \: z} 
\end{eqnarray}
\noindent where the $\beta_i$ are the fractions of the $p^\prime$
momentum carried by the various particles $i$ (see (\ref{eq:z1}))
and where for simplicity we neglect $\kappa_T$ in the quark loop.

Thus, we see that $|x^\prime| \ll x$.  The entire\footnote{Recall
that spin-one $t$ channel gluon exchange leads to an energy
independent (that is $x$ independent) cross section.} origin of
the $x$ dependence of Fig.\ 4 comes from the integral over the
rapidity of the $s$-channel gluon
\begin{equation}
\int_x^1 \; \frac{d x_s}{x_s}
\label{eq:d10}
\end{equation}
\noindent where the limit\footnote{In general the limit is $x_s >
\max \{x, x^\prime\}$, but in our case $x \gg |x^\prime|$.} comes
from $x_s = x_1 - x \simeq x_1 > x$.   This is the origin of the
BFKL $\ln (1/x)$ contribution\footnote{The virtual diagrams which
lead to the Reggeisation of the $t$ channel gluons also have the
same $\ln (1/x)$ structure.}.  Diagrams with further gluon rungs
give $\alpha_S^n \ln^n (1/x)$ contributions which are all
resummed by the BFKL equation for unintegrated gluon density
$f_{\rm BFKL} (x, k_T^2)$, with
$$
x \; \approx \; (Q^2 + M_\psi^2)/W^2 \; = \; 4
\overline{Q}^2/W^2.
$$
\indent To relate $f_{\rm BFKL}$ to the conventional gluon
density, which
satisfies GLAP evolution, we must integrate over $k_T^2$
\begin{equation}
xg (x, Q^2) \; = \; \int^{Q^2} \; \frac{dk_T^2}{k_T^2} \; f_{\rm
BFKL} \: (x, k_T^2),
\label{eq:e10}
\end{equation}
\noindent and so the $\gamma_T^{(*)} p \rightarrow J/\psi p$
forward amplitude (\ref{eq:a5}) becomes
\begin{equation}
A \; = \; i 4 \pi^2 M_\psi \: e_c g_J \: \alpha_S \; \int_0^\infty \;
\frac{dk_T^2}{2 \overline{Q}^2 (\overline{Q}^2 + k_T^2)} \;
\frac{\partial (x g (x, k_T^2))}{\partial k_T^2}.
\label{eq:f10}
\end{equation}
\noindent The integral converges for $k_T^2 \gg \overline{Q}^2$,
but the gluon distribution $xg (x, k_T^2)$ is not known as $k_T^2
\rightarrow 0$.  We therefore write
\begin{equation}
A \; \simeq \; i 2 \pi^2 \: M_\psi \: e_c g_J \: \alpha_S \;
\left [ \frac{xg (x, Q_0^2)}{\overline{Q}^4} \; + \;
\int_{Q_0^2}^\infty \; \frac{dk_T^2}{\overline{Q}^2 \:
(\overline{Q}^2 + k_T^2)} \; \frac{\partial xg (x,
k_T^2)}{\partial k_T^2} \right ],
\label{eq:g10}
\end{equation}
\noindent where in the first term in the brackets we have
neglected the effects of $k_T^2$ in comparison with
$\overline{Q}^2$.  Eq.\ (\ref{eq:g10}) allows the effects of the
gluon $k_T$ to be estimated. 
Eq.\ (18) is true to ${\cal O} (Q_0^2/\overline{Q}^2)$; we
investigate below the sensitivity of our results to variation of
the choice of $Q_0^2$.
The modification of (\ref{eq:a2})
becomes apparent when we use (\ref{eq:g10}) to calculate the
forward production cross section.  We have
\begin{equation}
\left . \frac{d \sigma}{dt} \; (\gamma^* p \rightarrow \psi p)
\right |_0 \; = \; \frac{\Gamma_{ee} \: M_\psi^3 \: \pi^3}{48
\alpha} \; \alpha_S (\overline{Q}^2)^2 \; \biggl [ \ldots\ldots
\biggr ]^2 \; \left (1 + \frac{Q^2}{M_\psi^2} \right ),
\label{eq:h10}
\end{equation}
where the $[\ldots]$ contain the entry shown in brackets in 
(\ref{eq:g10}).
In fact the accuracy of formula (19) is
even better than the BFKL approximation since the main part of
the corrections may be hidden inside the experimentally
determined values of $g (x, Q^2)$ and $\Gamma_{ee}$.  To be more
precise, if we neglect the $t_{min}$ effects of section 2(b),
eq.\ (17) is valid for a gluon distribution obtained from any
evolution equation.

\medskip
\medskip
\noindent {\bf (b)  Discussion of $t_{\rm min}$ effects} \\

We have tacitly assumed that we are considering a forward \lq\lq
elastic" scattering amplitude with $t = 0$.  However, for $\gamma
p \rightarrow J/\psi p$ the minimum value of $|t|$ is
\begin{equation}
t_{\rm min} \; = \; \left ( \frac{Q^2 + M_\psi^2}{W^2} \: m_p
\right )^2 \; \simeq \; x^2 m_p^2.
\label{eq:i10}
\end{equation}
\noindent This result is evident from (\ref{eq:b10}); we have to
transfer longitudinal momentum through the $t$-channel two-gluon
exchange.  We have already checked that the difference between
the momentum fractions $x$ and $x^\prime$ of the two gluons does
not affect the BFKL identification (\ref{eq:a8}) of $G$; it
contributes beyond the leading $\ln (1/x)$ approximation.  For
GLAP we could, in principle, recalculate Fig.\ 4 with
Altarelli-Parisi kernels \cite{glr} which take into account the
difference $x \neq x^\prime$.  Alternatively we can estimate the
effect from the analytic structure of $A$ in the complex $t$
plane, where the nearest singularity is the $2 \pi$ threshold at
$t_0 = 4m_\pi^2$.  Extrapolation from $t_{\rm min}$ to $t = 0$,
where the identification $xg (x, \overline{Q}^2)$ is true, gives
at most a correction of order $t_{\rm min}/t_0$.  In fact the $2
\pi$ singularity is weak, and $t_0 \sim m_\rho^2$ is a more
representative value \cite{ag}.  In any case we expect the
$t_{\rm min}$ effects to be small for $x < 0.1$.

\medskip
\medskip
\noindent {\bf (c)  Relativistic effects in the $J/\psi$ wave
function}\\

Here we discuss the effect of the motion of the quarks within the
$J/\psi$.  This Fermi motion has been considered in ref.\
\cite{fks} where it was concluded that it will lead to a
significant suppression of the cross section for diffractive
production.  Only longitudinal vector meson production was
studied there, whereas for photoproduction we have to deal with
transversely polarised $J/\psi$ mesons.

To estimate these relativistic effects we combine knowledge of
the $J/\psi$ wave function $\psi^J (\kappa_T, z)$ with the
structure of the coupling of the photon to the $c\overline{c}$
pair.  For transverse and longitudinally polarised photons the
amplitudes are of the form
\footnote{Here we neglect $k_T$ in comparison with $\kappa_T$.  It
is straightforward to derive formulae in the presence
of both $k_T$ and $\kappa_T$.  To simplify the presentation we
choose to investigate these corrections in turn and hence impose
the leading $\log Q^2$ approximation in eqs.\ (21) and (22)
[1,2].  Correlated effects will occur, but at the level of other
uncertainties.}$^,$\footnote{Eqs.\ (21) and (22) are derived assuming 
the form of the $J/\psi$ light cone wave function given in (2.22) of 
ref.\ [2].  The derivation is subtle and will be presented elsewhere
in a more general study of diffractive 
vector meson production\cite{lmrr}.}

\begin{eqnarray}
\label{eq:a11}
A (\gamma^T p \rightarrow J/\psi p) 
\propto  \frac{1}{m_c}  \int  \frac{2 [z^2 + (1 - z)^2] \kappa_T^2 
\tilde{Q}^2 + m_c^2 (\tilde{Q}^2 - \kappa_T^2)}{(\tilde{Q}^2 + \kappa_T^2)^3} 
\: xg (x, \tilde{Q}^2+ \kappa_T^2) & \psi^J (\kappa_T, z) \nonumber \\ 
 &  \times \: d^2 \kappa_T \:dz \nonumber \\
\end{eqnarray}
\vspace{-0.75cm}
\begin{eqnarray}
\label{eq:a12}
A (\gamma^L p \rightarrow J/\psi p) & \propto & \int \; \frac{2z (1 - z) \: 
\sqrt{Q^2} \: (\tilde{Q}^2 - \kappa_T^2)}{(\tilde{Q}^2 + \kappa_T^2)^3} \; xg 
(x, \tilde{Q}^2 + \kappa_T^2) 
 \psi^J (\kappa_T, z) \: d^2 \kappa_T \: dz 
\end{eqnarray}
\noindent where $\tilde{Q}^2 = z (1 - z) Q^2 + m_c^2$
and $x=(Q^2+M_{\psi}^2)/W^2$.  These expressions may 
be compared with the simplified non-relativistic approximation
in which we take $z = \frac{1}{2}$ and $\kappa_T = 0$, see
(\ref{eq:a3}). 
 Thus, if we have knowledge of the $J/\psi$ wave
function we can compute the correction factor $F^2$ arising from
the use of a more realistic $J/\psi$ wave function.  In
particular, for $J/\psi$ photoproduction we have from
(\ref{eq:a11})
\begin{equation}
F^2 \; = \; \left | \frac{\int \: (1 + v_T^2)^{-3} \: xg (x, m_c^2 (1 + v_T^2))
 \: \psi^J (\kappa_T, z) \: dv_T^2}{xg (x, m_c^2) \: \int \: \psi^J (\kappa_T, 
 z) \: dv_T^2} \right |^2 
\label{eq:a13}
\end{equation}
\noindent with $v_T^2 = \frac{2}{3} v^2 = \kappa_T^2/m_c^2$
where $v$ is the velocity of the quarks in the $J/\psi$ rest
frame.  In writing (23) we have set $z = \frac{1}{2}$ in the numerator, 
consistent with the approximation of neglecting contributions of ${\cal O} 
(v_T^4)$.

If we take a Gaussian form for the $J/\psi$
wave function, $\psi^J = A \exp (-a \: \kappa_T^2/m_c^2)$, then
$\langle v_T^2 \rangle = \frac{2}{3} \langle v^2 \rangle = 1/2a$.
Estimates of $\langle v^2 \rangle$ obtained from
studies of charmonium \cite{vv} vary in the range $0.12
\lapproxeq \langle v^2 \rangle \lapproxeq 0.25$,
with for example a recent lattice calculation \cite{lat} giving
$\langle v^2 \rangle$ values in the interval 0.18 to 0.12 as
$m_c$ increases from 1.45 to 1.85 GeV.  If we evaluate (23) using 
these ranges of $\langle v^2 \rangle$ and $m_c$ for each of the three 
recent sets of partons [GRV, MRS(A$^\prime$, G)] then we find that 
the suppression $F^2$  of the cross
section for diffractive $J/\psi$ photoproduction at HERA lies in the
interval
\begin{equation}
0.4 \; \lapproxeq \; F^2 \; \lapproxeq \; 0.6.
\label{eq:a14}
\end{equation}
The suppression due to the factor $(1 + v_T^2)^{-3}$ is partly 
compensated by the larger scale at which the gluon is evaluated.  
An independent study of
the relativistic corrections to $J/\psi$ photoproduction has been
made by Jung et al.\ \cite{jung}.  Using the gluonic and leptonic
widths of the $J/\psi$ they estimate $\langle v^2 \rangle = 0.16$
with $m_c = 1.43$ GeV, values which are quite compatible
with the above estimates.  The correction factor $F^2$ evaluated 
using these values of $\langle v^2 \rangle$ and $m_c$ is shown in 
Table 1 over the range of $x$ relevant to diffractive $J/\psi$ 
photoproduction at HERA.

\begin{table}[htbp]
\caption{The correction factor $F^2$ of (23) evaluated using three 
recent sets of partons.}
\begin{center}
\begin{tabular}{|c|c|c|c|} \hline
$x$ & $F^2$ (GRV) & $F^2$ (A$^\prime$) & $F^2$ (G) \\ \hline
$5 \times 10^{-4}$ & 0.46 & 0.54 & 0.48 \\
$10^{-3}$ & 0.45 & 0.52 & 0.47 \\
$2 \times 10^{-3}$ & 0.45 & 0.50 & 0.47 \\
$5 \times 10^{-3}$ & 0.44 & 0.48 & 0.46 \\ \hline
\end{tabular}
\end{center}
\end{table}

Besides the correction factor coming from the $J/\psi$ wave
function weighted by the $\gamma \rightarrow c\overline{c}$
coupling, we have to consider the effect arising from the
inequality $2 m_c \neq M_\psi$.  At first sight such a kinematic
effect might appear to be negligible.  
However we note that eq.\
(2) is written in terms of $M_\psi^2$, while in eqs.\ (21) and
(22) the current quark mass $m_c$ enters.  Thus we see that eq.\
(2) comes from eqs.\ (21) and (22), in the small $\kappa_T$
limit, with $m_c^2$ replaced by $\frac{1}{4} M_\psi^2$.
The other factors, such as the
leptonic width and the photon propagator, depend directly on
$M_\psi$.  So for photoproduction, with $Q^2 = 0$, the total
relativistic correction factor is estimated to be
\begin{equation}
F^2 \; \left ( \frac{M_\psi}{2 m_c} \right )^8 \; \approx \; 1
\label{eq:b14}
\end{equation}
\noindent with a large uncertainty of at least $\pm 30\%$.
Thus it turns out that the
best prescription is as follows.  We should replace $m_c^2$ by
$\frac{1}{4} M_\psi^2$ in eqs.\ (21) and (22) so that the mass
factor disappears from eq.\ (25) and the correction factor $F^2$
becomes closer to 1.  Of course this does not decrease the
overall uncertainty.
However we
emphasize that the uncertainty mainly affects the normalization of the
gluon density extracted from $J/\psi$ diffractive data, 
and that the prediction for the $x$ dependence should be much more reliable.
We see from Table 1 that the gluon density leads to a small
$x$ dependence of $F^2$, which we neglect in our comparison
with the data below, but which should
be included when the $J/\psi$ data become more precise.

We see from (\ref{eq:a11}) and (\ref{eq:a12}) that the
suppression $F^2$, arising from the use of the relativistic wave
function, decreases with increasing $Q^2$ due to the presence of
the $z (1 - z) Q^2$ term.  In fact since
\begin{equation}
z (1 - z) \; \approx \; \textstyle \frac{1}{4} \: (1 - \langle
v_z^2 \rangle ),
\label{eq:a15}
\end{equation}
\noindent where $\langle v_z^2 \rangle = \frac{1}{3} \langle v^2
\rangle$, we even expect, as $Q^2$ increases, that eventually
the factor $F^2$ will give an enhancement, {\it not} a
suppression, of the cross section.  When we take Fermi motion 
into account the ratio of diffractive
$J/\psi$ production from longitudinal and transverse photons
becomes
\begin{equation}
\frac{\sigma_L}{\sigma_T} \; \approx \; \frac{Q^2}{4m_c^2} \:
(1 - \langle v^2 \rangle )^2.
\label{eq:l6}
\end{equation}

Our result disagrees with the conclusions of ref.\cite{fks}
where a large suppression (of at least a factor of 3) was found
to arise from Fermi motion.  There are several differences in our
calculations.  First ref.\cite{fks} uses a potential model $J/\psi$
wave function whereas here we employ a Gaussian form.  Our
considered range of $\langle v^2 \rangle$ embraces a large class
of wave functions and covers the results obtained using the wave
function of any reasonable charmonium model.  (Incidentally it is
worth mentioning that the recent lattice calculation\cite{lat} gives a
wave function better described by a Gaussian form than by the
more
sophisticated potential models).  However, most of the difference
between our results and that of ref.\cite{fks} has a different
origin.  Part of the difference is due to the inclusion of the
mass factor in (25).  The remaining discrepancy arises from the
use of the appropriate wave function of the virtual photon and 
the use of the correct scale for the gluon in eq.\ (21).

\medskip
\medskip
\noindent {\bf (d)  Rescattering or absorption of $c\overline{c}$
quark-pair}\\

The shadowing effects of the gluons are already included in $xg
(x, \overline{Q}^2)$ if the distribution is determined in a
global parton analysis of experimental data.  Here we are
concerned with the rescattering or absorption of the
$c\overline{c}$ pair as it transverses through the proton.  A
typical diagram is shown in Fig.\ 5.  Let us study the first
shadowing correction (with $n = 1$) arising from the exchange of
one extra pair of gluons with transverse momenta $\pm \vec{K}_T$,
where the Sudakov decomposition (see (\ref{eq:z1})) of the gluon
4-momentum is
\begin{figure}[htb]
\vspace{4cm}
\includegraphics{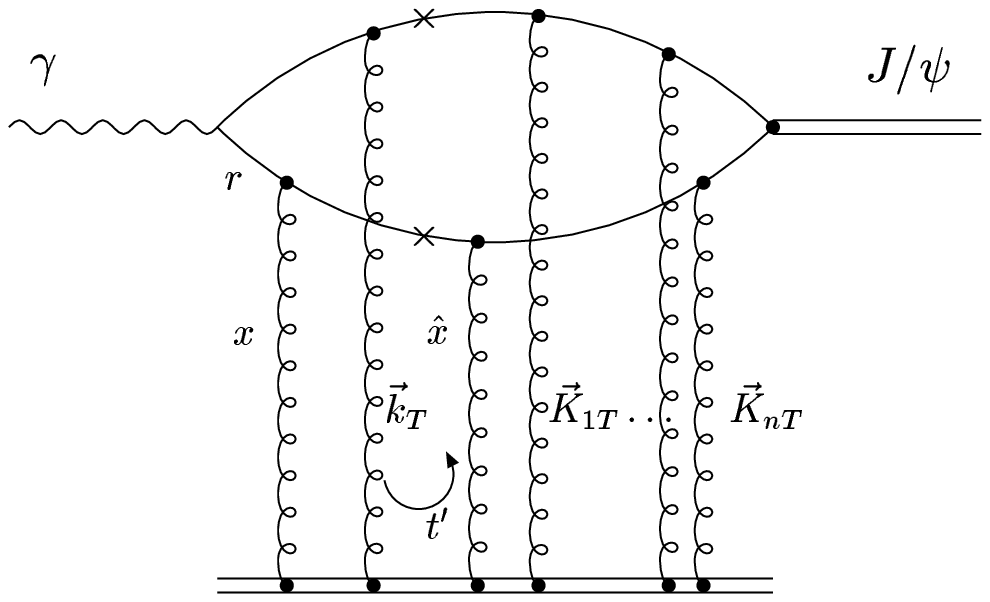}
\vspace{1cm}
\caption{ {\normalsize  A typical diagram leading to a
$c\overline{c}$
rescattering correction.  In the text we compute the $n = 1$
contribution.  All quark lines are on-shell except for that with
momentum $r$, as in Fig.\ 3.}}
\label{fig:5}
\end{figure}

\begin{equation}
K_1 \; \equiv \; K \; = \; \alpha_K p^\prime \: + \: \beta_K
q^\prime \: + \: \vec{K}_T.
\label{eq:a16}
\end{equation}
\noindent To compute the extra gluon loop we first close the
contour of the integrations over $\alpha_K$ and $\beta_K$ on the
quark propagators marked by small crosses on Fig.\ 5.  Hence we
again have all quarks on-mass-shell, except for that with
momentum $r$.  Thus we do not change the result of the
integration over the $c\overline{c}$ loop.  If we take into
account the four different ways the extra gluons can couple to
the $c$ and $\overline{c}$, then we have to include a total of 16
diagrams, and the integration over $\vec{k}_T$ and $\vec{K}_T$
gives, in analogy with (\ref{eq:a5}), an amplitude of the form
\begin{eqnarray}
\int \: \frac{d^2 k_T}{k_T^4} \: \int \: \frac{d^2 K_T}{K_T^4} \:
\; \frac{1}{2} \: \biggl [ \frac{2}{\overline{Q}^2} & - &
\frac{2}{\overline{Q}^2 + k_T^2} \: - \: \frac{2}{\overline{Q}^2
+ K_T^2} \: + \: \frac{1}{\overline{Q}^2 + |\vec{k}_T +
\vec{K}_T|^2} \nonumber \\
& + & \frac{1}{\overline{Q}^2 + |\vec{k}_T - \vec{K}_T|^2} \biggr
] \; f_{\rm BFKL} (x, k_T^2) \: f_{\rm BFKL}
(\hat{x}, K_T^2).
\label{eq:a17}
\end{eqnarray}
\noindent We again neglect $\kappa_T$.  The non-zero value of
$\hat{x}$ arises because we have to put a quark on-shell with an
additional momentum $\vec{K}_T$.  We therefore need to transfer
to the quark a proton momentum fraction
\begin{equation}
\hat{x} \; = \; \frac{K_T^2}{z W^2} \; \simeq \;
\frac{\overline{Q}^2}{2 z W^2} \; \simeq \; \frac{x}{4},
\label{eq:a21}
\end{equation}
\noindent since typically we have $K_T^2 \sim \overline{Q}^2/2$. 
For $k_T, K_T \ll \overline{Q}$, after neglecting the angular
correlation between $\vec{k}_T$ and $\vec{K}_T$, that is setting
$$
\langle (\vec{k}_T . \vec{K}_T)^2 \rangle \; = \; \frac{1}{2} \:
k_T^2 \: K_T^2,
$$
\noindent the expression in square brackets in (\ref{eq:a17})
becomes
$$
\frac{1}{2} \; \biggl [ \ldots.. \biggr ] \; = \;
\frac{k_T^2}{\overline{Q}^4} \; \frac{4 K_T^2}{\overline{Q}^2}.
$$
\indent We expand the integrand in (\ref{eq:a17}) assuming that
$k_T^2, K_T^2 \ll \overline{Q}^2$.  If we keep the leading terms then we
find
\begin{equation}
\frac{\Omega}{2} \; = \; \frac{\pi \alpha_S (\overline{Q}^2)}{3
b} \; \int \; \frac{d K_T^2}{\overline{Q}^2 + K_T^2} \;
\frac{\partial (\hat{x} g (\hat{x}, K_T^2))}{\partial K_T^2},
\label{eq:a18}
\end{equation}
\noindent where $\Omega$ is defined so that the correction to the
amplitude is
\begin{equation}
A \; = \; A_0 (1 - \textstyle \frac{1}{2} \Omega).
\label{eq:a19}
\end{equation}
\noindent As the correction is not too large we can eikonalize
(\ref{eq:a18}) by hand, directly in the momentum representation,
and write the final expression in the form
\begin{equation}
\sigma = \sigma_0 \exp (- \Omega).
\label{eq:a199}
\end{equation}
\noindent The factor $1/b$ in (\ref{eq:a18}) comes from the
integral over the \lq\lq reggeon" loop in Fig.\ 5
\begin{equation}
\int  dt^\prime \: e^{- bt^\prime} \; = \; 1/b,
\label{eq:a20}
\end{equation}
\noindent where we use the experimental slope $b = 4.5$
GeV$^{-2}$.  The $K_T^2$ integration in (\ref{eq:a18}) is
evaluated just as in (\ref{eq:f10}) and (\ref{eq:g10}).

It is worthwhile mentioning that the simple estimate of
(\ref{eq:a18}) is in good agreement with the more detailed
treatment of ref.\ \cite{glm}.

\medskip
\medskip
\noindent {\bf (e)  Next-to-leading order corrections}\\

There are expectations that higher-order radiative corrections
will have a negligible effect.  We refer to four possible types
of radiation shown all together in Fig.\ 6. 

The contribution of the diagram with the single extra gluon
denoted by (a) is already incorporated in $\Gamma (J/\psi
\rightarrow e^+ e^-)$, while (b) is hidden in $\alpha_S$ and (c)
is subsumed in $xg (x, Q^2)$.  Only correction (d) survives, but
this is suppressed by a factor $\tau_{\rm int}/\tau_f$, which is
small for high $W$, that is for $x \ll 1$, see the comment in the
caption to Fig.\ 1.
\begin{figure}[htb]
\vspace{6cm}
\includegraphics{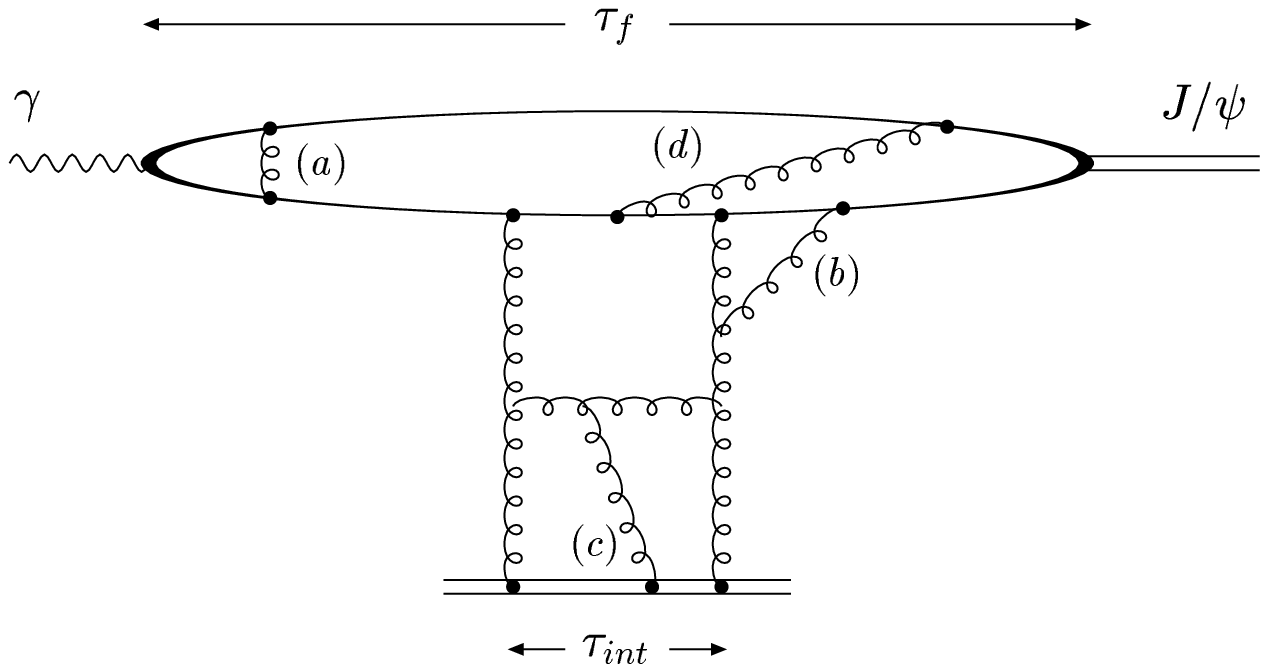}
\vspace{0.25cm}
\caption{ {\normalsize   Four different types of radiative
correction,
denoted by (a,b,c,d), are shown together on the lowest-order
diagram for diffractive $J/\psi$ production.}}
\label{fig:6}
\end{figure}

Independent evidence for small radiative effects comes from a
study of next-to-leading order corrections in {\it inelastic}
$J/\psi$ production at HERA \cite{zerwas}.  The lowest-order
subprocess is shown in Fig.\ 7.  It is found that the corrections
to inelastic $J/\psi$ production are in general appreciable
($\gapproxeq 50\%$).  However, they reduce almost to zero when
$\hat{s} \rightarrow M_\psi^2$, that is when the emitted gluon
$k$ is very soft, which corresponds to our pomeron with
$|x^\prime| \ll x$, see (\ref{eq:c10}).
\begin{figure}[htb]
\vspace{6cm}
\includegraphics{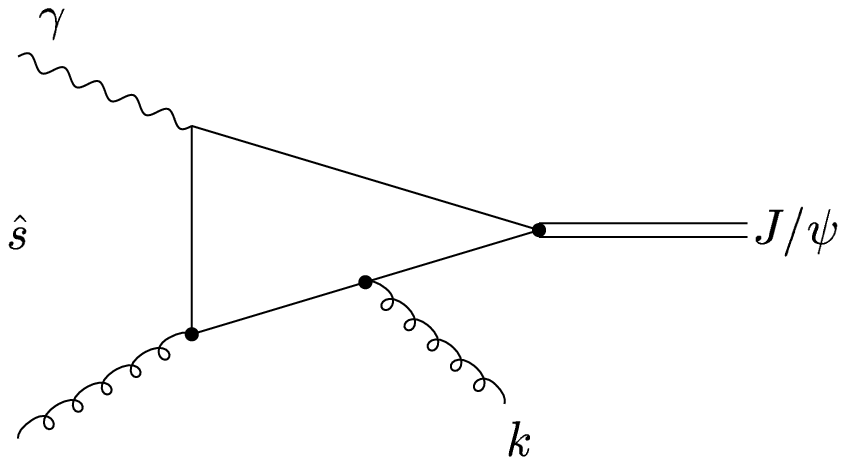}
\caption{ {\normalsize  The lowest-order subprocess, $\gamma g
\rightarrow J/\psi g$, describing inelastic $J/\psi$ production. 
The emitted gluon $k$ is necessary to satisfy the colour and spin
constraints of the $J/\psi$. }}
\label{fig:7}
\end{figure}

\medskip
\medskip
\noindent {\bf (f)  Inclusion of the real part}\\

So far we have calculated only the imaginary part of the
amplitude.  At high energy $W$, that is at small $x$, our
positive-signature amplitude behaves as
$$
A \; \propto \; x^{- \lambda} \: + \: (-x)^{- \lambda}.
$$
\noindent Providing $\lambda$ is small, the amplitude is
dominantly imaginary and the real part can be calculated as a
perturbation
\begin{equation}
\frac{{\rm Re} A}{{\rm Im} A} \; \approx \; \frac{\pi}{2} \lambda
\; \approx \; \frac{\pi}{2} \: \frac{\partial \ln A}{\partial \ln
(1/x)} \; \approx \; \frac{\pi}{2} \: \frac{\partial \ln (xg (x,
\overline{Q}^2))}{\partial \ln (1/x)}.
\label{eq:a22}
\end{equation}

\medskip
\medskip
\noindent {\large \bf 3.  Comparison with diffractive $J/\psi$
photoproduction data} \\

We use the perturbative QCD formula (\ref{eq:h10}) together with
the
corrections which we detailed in section 2, to calculate the
cross section for diffractive $J/\psi$ photoproduction
$$
\sigma (\gamma p \rightarrow J/\psi \: p) \; = \; \left .
\frac{1}{b} \; \frac{d \sigma}{dt} \; (\gamma p \rightarrow
J/\psi \: p) \right |_0
$$
\noindent as a function of the $\gamma p$ centre-of-mass energy
$W$.  We take the experimental value for the slope parameter:  $b
= 4.5$ GeV$^{-2}$.  The prediction depends on the square
of the gluon density, $[xg (x, \overline{Q}^2)]^2$, at $x =
M_\psi^2/W^2$ and for values of $\overline{Q}^2$ in the region of
$M_\psi^2$.  For illustration we calculate the $W$-dependence of
the cross section using the gluon distributions of three of the
latest sets of partons, namely GRV \cite{grv} and MRS(A$^\prime$,
G) \cite{mrs}.  The latter partons are extrapolated below $Q^2 =
4$ GeV$^2$ with next-to-leading GLAP evolution.

\begin{figure}[htb]
\includegraphics{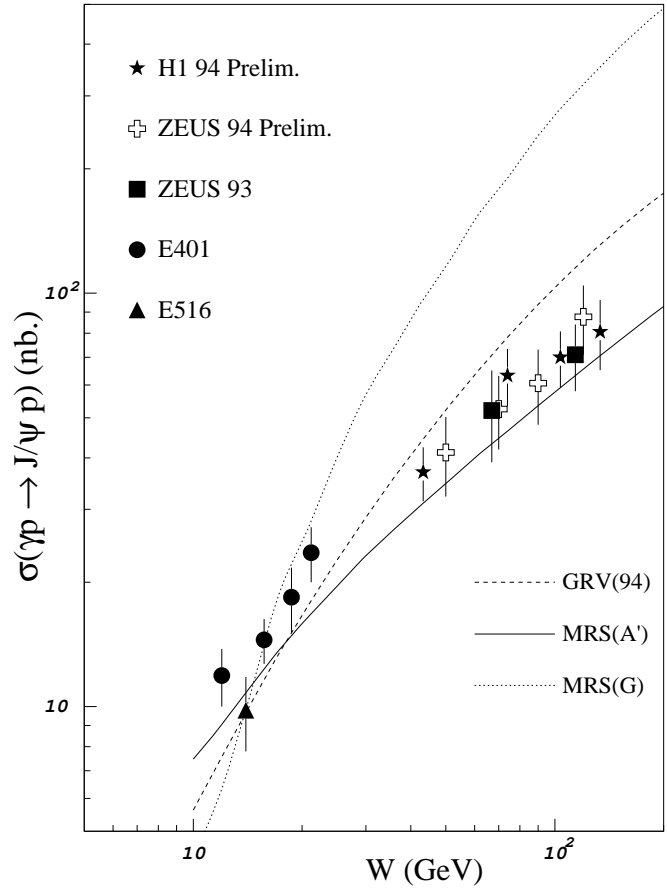}
\vspace{12.5cm} 
\caption{ {\normalsize  The measurements [5,6] of the
cross section for diffractive $J/\psi$ photoproduction compared
with the full perturbative QCD prediction, as described in
section 2, obtained from the three latest sets of partons.}}
\label{fig:8}
\end{figure}

It could be argued that it is inappropriate to use
GLAP-based gluons in our resummed $\log (1/x)$ formalism. 
However,
they should be regarded simply as a parametrization of the data. 
The non-perturbative parameters are a good representation of all
physical effects such as BFKL smearing and shadowing corrections.
It is well known that the present data cannot distinguish the
underlying perturbative physics.  However, data do distinguish
between the parametrizations.

The three
predictions are compared with the available high energy $J/\psi$
photoproduction data in Fig.\ 8. 
 The curves correspond to the
choice of lower limit $Q_0^2 = 1$ GeV$^2$ in the integral
over the gluon $k_T^2$ in (\ref{eq:h10}).  We explored the
sensitivity
of the predictions to variation of $Q_0^2$ over the range 0.5 to
2 GeV$^2$ for the GRV gluon, and 1 to 4 GeV$^2$ for the MRS
gluons.  Depending on the parton set, the cross section values
change by $\pm 15\%$ at $W = 10$ GeV, increasing to $\pm 25\%$
at $W = 100$ GeV as $Q_0^2$ is varied.

In Table 2 we quantify the effect of sequentially improving the
prediction of the lowest-order formula (\ref{eq:a2}), first to
(\ref{eq:h10}), which goes beyond the leading $\ln Q^2$
approximation
to include the effects of the transverse momenta of the exchanged
gluons, and then to include the factor $\exp (- \Omega)$, with
$\Omega$ given by (\ref{eq:a18}), to allow for the
$c\overline{c}$
rescattering on the proton.  Table 2 shows the size of the
effects at $W = 100$ GeV.  We see that the two corrections to the
lowest-order formula partially compensate each other.  The
corrections decrease with decreasing energy $W$, and are found to
be insignificant for $W \lapproxeq 20$ GeV.

\begin{center}
Table 2: The cross section for diffractive $J/\psi$
photoproduction
\end{center}
\begin{center}
\begin{tabular}{|c|c|c|c|} \hline
& \multicolumn{3}{|c|}{$\sigma (\gamma p \rightarrow J/\psi \:
p)$
in
$nb$ at $W = 100$ GeV} \\ \cline{2-4}
\raisebox{1.5ex}[0pt]{Partons} & lowest-order & + gluons $k_T$ &
+ $c\overline{c}$ rescatt. \\ \hline
GRV & 128 & 157 & 88 \\
MRS (A$^\prime$) & 47 & 79 & 55 \\
MRS (G) & 262 & 331 & 208 \\ \hline
 \end{tabular}
\end{center}
\noindent In addition to the above effects, the predictions shown
in Fig.\ 8 also include the real part contribution via
(\ref{eq:a22}).  According to (\ref{eq:b14}) the $J/\psi$
relativistic effects are estimated to leave the cross section
unaltered, but to introduce a sizeable normalization uncertainty.

\medskip
\medskip
\noindent {\large \bf 4.  Conclusions}\\

\begin{figure}[htb]
\includegraphics{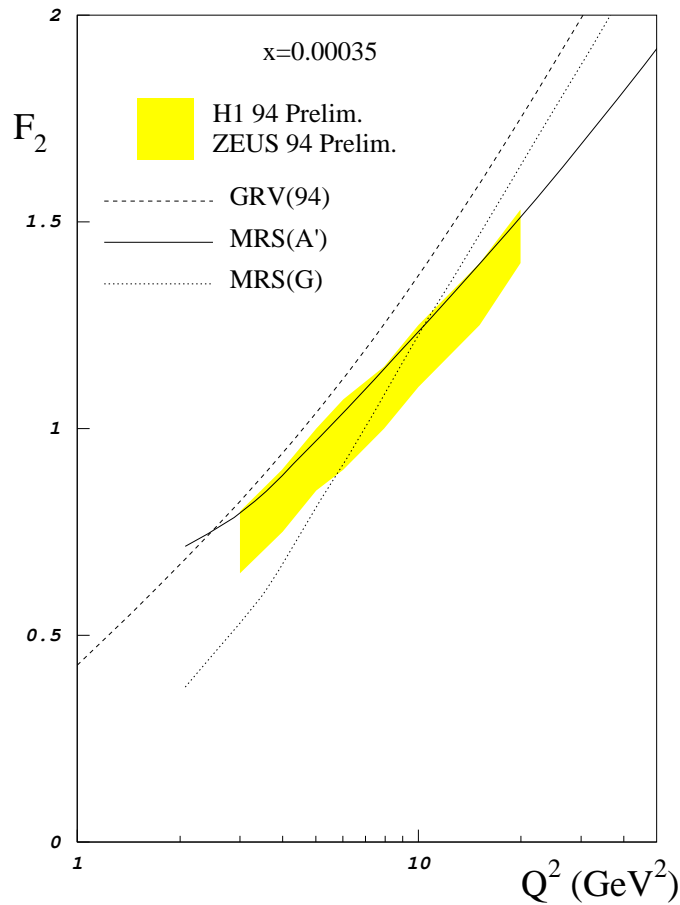}
\vspace{12.5cm} 
\caption{ {\normalsize  The values of $F_2$ versus $\ln Q^2$ at
$x = 3.5 \times 10^{- 4}$ obtained from GRV [7] and
MRS(A$^\prime$,G) [8] partons.  The shaded band delimits
the most recent measurements of the structure function obtained
by the H1 and ZEUS collaborations [18].}}
\label{fig:9}
\end{figure}

We conclude that the HERA $J/\psi$ photoproduction data offer a
unique opportunity to distinguish between the various gluon
distributions in a kinematic region $(x = M_\psi^2/W^2 \simeq
10^{-3}$ and $\overline{Q}^2 \simeq 2.5$ GeV$^2)$ where they
are particularly distinct.  In order to make a meaningful
comparison we have computed several corrections to the
lowest-order formula for the $J/\psi$ cross section.  The
uncertainties are found to be greater in the predicted size of
the cross section than in the $W$ or, equivalently, the $x$
dependence.  That is the shape, rather than the normalisation, is
the better discriminator between the various gluons.  
The comparison of the $J/\psi$ cross section data
with the values calculated using the gluons of three of the
latest parton sets was shown in Fig.\ 8.  The shape 
(and the normalisation) predicted by
the MRS(A$^\prime$) gluon is, within the theoretical accuracy,
in excellent agreement with the
measured values of the $J/\psi$ photoproduction cross section. 
In fact the data appear to favour the MRS(A$^\prime$) gluon over
the GRV gluon, and to rule out that of MRS(G).

A similar preference for the MRS(A$^\prime$) gluon is found by
inspecting the slope of the most recent HERA measurements of
the structure function $F_2$ as a function of ln$Q^2$ as shown,
for example, in Fig.\ 9.  
This slope is an independent measure of the gluon at small $x$,
though, as anticipated, it is not such a sensitive probe as the
$W$ dependence of the HERA diffractive $J/\psi$ cross section
data.  However, if combined together, the HERA measurements of
the diffractive cross section $\sigma (\gamma p \rightarrow
J/\psi p)$ and of the slope $\partial F_2/\partial \ln Q^2$ offer
an excellent opportunity to pin down the gluon density for $x
\lapproxeq 10^{-3}$.

\medskip
\noindent {\large \bf Acknowledgements}

We thank Drs.\ R.\ C.\ E.\ Devenish and M.\ R.\ Whalley for
organizing the Durham Workshop on \lq\lq HERA Physics", where
this work originated.  We thank the UK Particle Physics and
Astronomy Research Council and the Royal Society for financial
support.  Also EML thanks CNPq of Brazil for financial support.

\bigskip

\end{document}